\documentclass[twocolumn,amsmath,amssymb,floatfix]{revtex4}

\usepackage{bm}
\usepackage{amsmath}
\usepackage{epsf}
\newcommand{\beq}{\begin{equation}}
\newcommand{\eeq}{\end{equation}}
\newcommand{\beqa}{\begin{eqnarray}}
\newcommand{\eeqa}{\end{eqnarray}}
\newcommand{\data}[1]{{\bf d}_{#1}}
\newcommand{\proj}[1]{{\bf P}_{#1}}
\newcommand{\weight}[1]{{\bf R}_{#1}}

\newcommand{\esignal}[1]{\hat{\bf s}_{#1}}
\newcommand{\signal}[1]{{\bf s}_{#1}}
\newcommand{\iden}{{\bf I}}
\newcommand{\tot}{{\rm tot}}
\newcommand{\noise}[1]{{\bf n}_{#1}}

\newcommand{\csignal}[1]{{\bf S}_{#1}}

\newcommand{\bcsignal}[1]{\bar {\bf S}_{#1}}
\newcommand{\cnoise}[1]{{\bf N}_{#1}}
\newcommand{\diag}{{\rm diag}}
\newcommand{\corr}[1]{{\bf C}_{#1}}
\newcommand{\dela}{{\Delta}}
\newcommand{\dD}{{\delta\!D}}

\newcommand{\etal}{{\it et al. }}
\newcommand{\de}{{\rm DE}}
\newcommand{\fsky}{f_{\rm sky}}
\newcommand{\sifsky}{f_{\scriptscriptstyle \mathrm{sky}}^{\scriptscriptstyle -1/2}}
\newcommand{\rooteigen}{\lambda_{\scriptscriptstyle i}^{\scriptscriptstyle -1/2}}

\newcommand{\ApJL}{Astrophys. J Lett.}
\newcommand{\ApJ}{Astrophys. J}
\newcommand{\PRL}{Phys. Rev. Lett.}
\newcommand{\PRD}{Phys. Rev. D}
\newcommand{\MNRAS}{Mon. Not. Roy. Astr. Soc.}
\newcommand{\ARAA}{Ann. Rev. Astron. Astrophys.}
\newcommand{\AsAs}{Astron. Astrophys.}
\newcommand{\amp}{\& }
\newcommand{\aut}[2]{{#2.\ #1,}}
\newcommand{\laut}[2]{{#2.\ #1,}}
\newcommand{\refs}[6]{#2, {#3},  {#4} (#5).}
\newcommand{\mrefs}[6]{#2, {#3},  {#4} (#5);}
\newcommand{\urefs}[5]{#2, #3, #4 (#5).}
\newcommand{\murefs}[5]{#2, #3, #4 (#5);}
\newcommand{\mybib}[2]{\bibitem{#2}}
\begin{document}

%@twocolumnfalse\endcsname
\title{Dark Energy and Matter Evolution from
Lensing Tomography}
\author{Wayne Hu}
\affiliation{
{}Center for Cosmological Physics, Department of Astronomy and Astrophysics, 
and Enrico Fermi Institute, University of Chicago, Chicago IL 60637
}

\begin{abstract}
\baselineskip 11pt
Reconstructed from lensing tomography, the evolution of the dark matter density field 
in the well-understood linear regime can provide model-independent constraints on the
growth function of structure and the evolution of the dark energy density.  
We examine this potential in
the context that high-redshift cosmology has in the future
been fixed by CMB measurements.  We construct  sharp tests for
the existence of multiple dark matter components or a dark 
energy component that is not a cosmological constant. 
These functional constraints can be transformed into physically 
motivated model parameters. From the growth function, the fraction of the dark matter in 
a smooth component,  such as
a light neutrino, may be constrained  to a statistical precision of $\sigma(f) \approx 0.0006 \sifsky$
by a survey covering a fraction of sky $\fsky$ 
with redshift resolution $\Delta z=0.1$.  For the dark energy, a parameterization
in terms of the present energy density $\Omega_{\de}$, equation of state $w$ and its
redshift derivative $w'$, the constraints correspond to $\sigma(w)=0.016 \sifsky$
and a mildly degenerate combination of the other two parameters.  For a fixed $\Omega_{\de}$,
$\sigma(w') = 0.046 \sifsky$;  for $\Omega_{\de}$ marginalized $\sigma(w')=0.069 \sifsky$.
\end{abstract}
\maketitle

\section{Introduction}

The weak gravitational lensing of faint galaxies \cite{weak} provides the most
direct probe of mass distribution in the universe (e.g.~\cite{BarSch01}).    Moreover the evolution of
clustering in the mass distribution is arguably the best theoretically-grounded 
probe of the dark energy and dark matter \cite{deprobes}.
Although observations of weak lensing on large scales \cite{weaklss} are still in the discovery
phase \cite{weakdet}, future wide-field surveys have the potential to rival
the statistical precision and cosmological utility 
 of luminosity distance measures from supernova surveys
\cite{Hut02,Hu01c}.  Even in the context of a precisely-determined homogeneous cosmology,
lensing measurements are unique in that they probe the clustering properties
of the dark matter and energy.  These are fixed by the homogeneous cosmology only
under particular assumptions of the particle constituents (e.g. cold dark matter
and scalar field dark energy) \cite{CalDavSte98,Hu98}. 

Much of the critical cosmological information
 lies in the temporal or radial direction.
A potential obstacle for weak lensing  is that the observables are
inherently two-dimensional.  All of the matter along the line-of-sight to
a distant source contributes to the lensing.    
For a family of  cosmological models that is described by a handful of parameters,
this is not a serious drawback.  Lack of radial information is largely 
compensated by a large angular dynamic range and external cosmological 
information.

Given the lack of compelling models for the dark energy and controversies surrounding
the phenomenology of the dark matter on small scales, it is interesting to consider 
a more model-independent approach.   Indeed recent studies of alternate
parameterizations of the dark energy have revealed potential ambiguities in
the interpretation of luminosity distance 
measurements \cite{WanGar01,MaoBruMcMSte02,WelAlb02,Teg01,HutSta02}.
To address these issues with weak lensing,  recovery of the temporal dimension
becomes critical.   

With future surveys that possess source photometric redshift information, recovery of
the lost information is possible in principle through tomography.   Photometric redshift
techniques
are already being applied and tested on current lensing data \cite{Witetal01}.
The full two-point statistical information can be regained by cross-correlating the lensing observables
on all source redshift planes \cite{Hu99}.   This method utilizes both the angular
clustering and the temporal evolution of the density field but obscures the
nature and  hence
the model-dependence of the information. Additionally, the joint observables are survey dependent
and computationally cumbersome 
to analyze.  

In this paper, we instead
isolate the temporal information by applying recently developed techniques to
reconstruct the radial density field itself \cite{Tay01,HuKee02}.   We will further 
focus solely on the linear regime where predictions are well-understood.  Even 
utilizing only this theoretically-clean subset of information in the data, future
surveys can potentially provide interesting model-independent constraints on
the properties of the dark energy and matter.

The outline of the paper is as follows.  In \S \ref{sec:tomography}, we discuss
the method for reconstruction and statistical forecasts.  In \S \ref{sec:growth}, 
we study constraints on the growth function for a fixed homogeneous cosmology and
in \S \ref{sec:density} the dark energy density evolution assuming pure cold dark matter.
We discuss these results in \S \ref{sec:discussion}.

\section{Tomographic Reconstruction}
 \label{sec:tomography}

We begin in \S \ref{sec:algorithm}  by briefly reviewing the tomographic reconstruction of the
dark matter density field 
in a fixed background cosmology as studied in \cite{HuKee02}.  We then 
generalize to the case where
the cosmology and the density field must be jointly recovered from the data
in \S \ref{sec:uncertaindz} and review Fisher techniques for statistical forecasts in \S \ref{sec:forecast}.  In \S \ref{sec:fiducial}, we outline the fiducial
cosmology and survey parameters used for illustrative purposes in the following
sections.

\subsection{Known Homogeneous Cosmology}
\label{sec:algorithm}

We consider the data to be the lensing convergence $\kappa$ in 
an angular pixel discretized into bins of source redshift composed
into a data vector $\data{\kappa}$.  Weak lensing dictates that
the data is a linear projection of an underlying density field plus noise in the 
convergence measurement $\noise{\kappa}$
\begin{equation}
\data{\kappa} = \proj{\kappa \dela} \signal{\dela} + \noise{\kappa}\,,
\end{equation}
with
\begin{eqnarray}
[\proj{\kappa\dela}]_{ij} =
\begin{cases}
\frac{3}{2} H_0^2 \Omega_m \dD_j \frac{(D_{i+1} - D_j)D_j}{D_{i+1}}
   & \text{ $D_{i+1} > D_j$ } ,\\
 0 & \text{ $D_{i+1} \le D_j$ },
\end{cases}
\label{eqn:denproj}
\end{eqnarray}
where $D$ is the comoving distance in a flat universe 
\begin{equation}
D(z) = \int_0^z {dz' \over H(z')}\,,
\label{eqn:distance}
\end{equation}
with $H^2=8\pi G \rho_{\rm tot}/3$ defining the Hubble parameter and
$\dD_j$ is the width of bin $j$.   Here and throughout subscripts
on matrices are labels, whereas matrix elements are denoted as $[\,]_{ij}$.
Here $\dela =(\delta \rho/\rho)/a$ is the density fluctuation in the
bin with the growth rate in a matter-dominated universe scaled out.
Note that these
distances depend on the assumed cosmology.
We assume for now that $D(z)$ has been fixed by other observations,
e.g. future supernovae surveys, but relax this assumption in
the following section.

The minimum variance estimator of the underlying density field
is given by \cite{HuKee02}
\begin{equation}
\esignal{\Delta} = \weight{\Delta\kappa} \data{\kappa}\,,
\label{eqn:densityest}
\end{equation}
where the reconstruction matrix
\begin{eqnarray}
\weight{\Delta\kappa} &=&
%[\proj{\kappa\Delta}^t \cnoise{\kappa\kappa}^{-1} \proj{\kappa\Delta}]^{-1}
\cnoise{\Delta\Delta}
\proj{\kappa\Delta}^t \cnoise{\kappa\kappa}^{-1}\,.
\label{eqn:radialweight}
\end{eqnarray}
Here
\begin{equation}
\cnoise{\Delta\Delta} =
[\proj{\kappa\Delta}^t \cnoise{\kappa\kappa} \proj{\kappa\Delta}]^{-1}\,,
\label{eqn:noisedelta}
\end{equation}
is the noise covariance of the estimator.  Note that $\weight{\dela\kappa}
\proj{\kappa \dela} =\iden$ so that the estimator is unbiased.

The statistical properties of the recovered density field
contain cosmological information.  
The recovered density field is  an average of
the density fluctuation over a window 
(or mask) $W_i({\bf x})$
defined by the
angular pixel and redshift binning.
The signal covariance of these density averages 
\begin{eqnarray}
\left[\bcsignal{\Delta\Delta}\right]_{ij} 
&=& \frac{\phi_i}{\phi_0} \frac{\phi_j}{\phi_0}
\int \frac{d^3 k}{(2\pi)^3} W_i({\bf k}) W_j^*({\bf k}) 
P(k)\,,\nonumber
\label{eqn:variancesimp}
\end{eqnarray}
where $P(k)$ is the linear power spectrum  today, $\phi_i = D_{\rm grow}(z_i)/a_i$ is the
linear decay rate of the potential field, with $D_{\rm grow}$ the linear
growth rate of the density field, normalized so that
$\phi_i = 1$ in the matter dominated regime, and
$W_i({\bf k})$ are the Fourier transforms of the windows.  The two-point
statistical properties of the reconstruction therefore contain information
on the growth rate and underlying power spectrum of the density field.

\subsection{Unknown Homogeneous Cosmology}
\label{sec:uncertaindz}

The inversion of  Eqn.~(\ref{eqn:densityest}) requires
an assumption of a distance-redshift relation in $\proj{\kappa\Delta}$.  
If this relation is not fixed
by external constraints, then the reconstructed density field will be a biased
measure of the true density field.   In this case, both the distance-redshift and growth rate
must be fit to the data.   

If the assumed $D(z)$ is close to the true $D(z)$, then the reconstruction of the 
previous section still serves as a useful representation of the data.  
The reconstruction matrix $\weight{\Delta\kappa}$
employs a slightly incorrect assumption of the
projection matrix so that
$\weight{\Delta\kappa} \proj{\kappa\Delta} \ne \iden$.  
The statistical properties of the estimated density field are encapsulated
in the noise  matrix (\ref{eqn:noisedelta}), which remains unchanged,
and a new signal covariance matrix
\begin{equation}
\csignal{\Delta\Delta} = 
 \weight{\Delta\kappa} \proj{\kappa\Delta}\, \bcsignal{\Delta\Delta}\,
 \proj{\kappa\Delta}^t \weight{\Delta\kappa}^t\,.
\end{equation}
The two-point statistics now also contain information about the
distance-redshift relation $D(z)$.

\subsection{Statistical Forecasts}
\label{sec:forecast}

The two-point statistical information in the reconstruction
can be exposed through the familiar Fisher approach (e.g. 
\cite{TegTayHea97}).  
If the parameters that underly the two-point correlation are
given by a vector $\signal{p}$, the Fisher matrix is given by
\begin{equation}
[ {\bf F}_{pp}]_{ij} = {N_{\rm pix}\over 2} {\rm tr} 
[ 
 \corr{\Delta\Delta}^{-1} 
 \corr{\Delta\Delta,i}
 \corr{\Delta\Delta}^{-1} 
 \corr{\Delta\Delta,j} ]
\label{eqn:fisher}
\end{equation}
with the covariance matrix
\begin{equation}
\corr{\Delta\Delta} = \cnoise{\Delta\Delta} + \csignal{\Delta\Delta}\,.
\end{equation}
Here we have assumed that the convergence is measured in $N_{\rm pix}$
independent pixels.  We will often write this factor as the
total sky coverage in independent pixels
$N_{\rm pix} = 4\pi \fsky/A_{\rm pix}$ where $A_{\rm pix}$ is the
pixel area in steradians.  A complete treatment would track
the small correlations between neighboring pixels 
on contiguous patches of sky \cite{HuKee02}.

The inverse of the Fisher matrix ${\bf F}_{pp}^{-1}$ 
gives an estimate of the covariance matrix ${\bf C}_{pp}$
of the measured parameters
$\esignal{p}$.  Note that under a re-parameterization of the space,
the Fisher matrix transforms as a covariant tensor
\begin{equation}
{\bf F}_{\tilde p \tilde p} = {\bf J}_{p\tilde p}^t {\bf F}_{pp} {\bf J}_{p\tilde p}, \qquad
[{\bf J}_{p\tilde p}]_{ij} \equiv \frac{\partial p_i}{\partial \tilde p_j}\,,
\label{eqn:fishertrans}
\end{equation}
We will use this
fact to go from model-independent parameterizations of the underlying
functions to model-dependent ones.

\subsection{Fiducial Model and Survey}
\label{sec:fiducial}

We take as a fiducial cosmology a flat $\Omega_{\rm tot}=1$ universe
with $\Omega_c = 0.3$ in cold dark matter, $\Omega_b=0.05$ in baryons, 
$\Omega_{\de}=0.65$ in dark energy; an equation of state of the
dark energy of $w(z)=-1$ corresponding to a cosmological constant; a
dimensionless Hubble constant, $h=0.65$, 
scalar spectral index $n=1$, and amplitude of the initial
curvature power spectrum $\delta_\zeta = 4.8 \times 10^{-5}$
($\sigma_8=0.92$, see \cite{Hu01c} for specific definitions of
parameters).
Since tomographic reconstruction will mainly be useful for
next-generation surveys, it is reasonable
to assume that CMB experiments will by that time have determined many of
the underlying cosmological parameters to high accuracy.
For simplicity we will here assume that $\Omega_c h^2$, 
$\Omega_b h^2$, $n$, $\delta_\zeta$ and $\Omega_{\rm tot}$ 
are completely fixed to their fiducial values.  We will return 
to this point in \S \ref{sec:discussion}. These parameters fix the
shape and high redshift normalization of the potential 
power spectrum and so we will work in the context that
only the growth function and distance-redshift relation need
be determined through lensing.

For definiteness, we will take the fiducial survey to be
defined with circular pixels of area $1 \deg^{2}$ 
so that 
the reconstructed density field 
is in the linear regime \cite{JaiSel97}.  Larger pixels would
reduce the number of independent pixels and hence increase the
sample variance in the signal dominated regime.
In the figures that follow, we take $\fsky=0.1$ ($\sim$ 4000 deg$^2$) but note that
all lensing errors may be rescaled as $(\fsky/0.1)^{-1/2}$.
For the redshift binning, 
we take $\Delta z=0.1$ out to $z=3$. With this binning, the signal covariance
matrix ${\bf S}_{\Delta\Delta}$ is nearly diagonal and the statistics reduce
to the evolution of the density variance in bins.

We will assume a convergence
noise spectrum of the form 
\begin{equation}
\cnoise{\kappa\kappa} = \diag [\gamma_{\rm rms}^2/
N_{i}]\,,
\end{equation} 
as appropriate for random intrinsic galaxy ellipticities.
We take $\bar n = 3.6 \times 10^5$ gal. deg$^{-2}$ and $\gamma_{\rm rms}=0.3$
as an estimate of the usable galaxies and
the shear noise per galaxy measured
from a space-based platform (A.~Refregier, private communication)
and form the number of galaxies per bin $N_i$ from a redshift distribution
\cite{Kai98}
\begin{equation}
\frac{ dN }{d z} \propto \frac{dD}{dz} D \exp[-(D/D_*)^4]\,,
\end{equation}
where $D_*$ is set to reproduce a median redshift $z_{\rm med}=1$.
These fiducial survey specifications are chosen to represent the upper range
of the capabilities of surveys in the foreseeable future.  

\begin{figure}[tb]
\centerline{\epsfxsize=3.5truein\epsffile{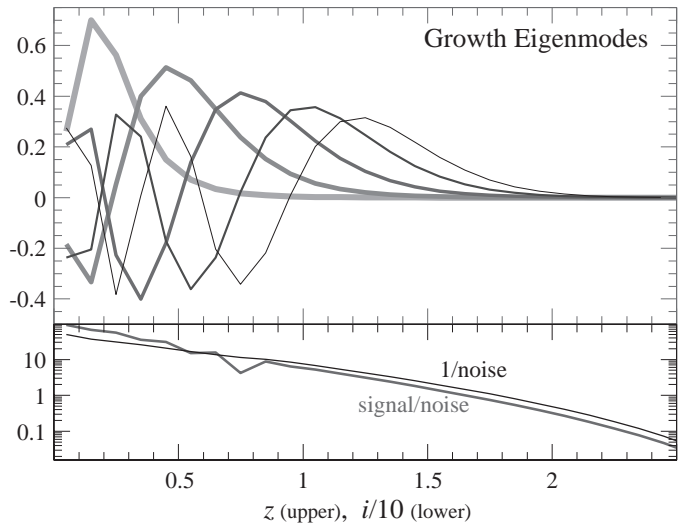}}
\caption{\footnotesize Principle components of the growth function: (upper) first five
eigenfunctions of the growth function (thick to thin) plotted in discrete redshift bins;
 (lower) rank ordered inverse rms noise $\rooteigen$ (thin) and 
signal-to-noise ratio.  Here and in the following figures the fiducial
model and survey are assumed throughout with $\fsky=0.1$ and $\Delta z=0.1$ redshift bins.}
\label{fig:eigengrow}
\end{figure}

\section{Growth Rate}
\label{sec:growth}

For a fixed distance-redshift relation and high redshift power spectrum,
the remaining degrees of
freedom in the two-point statistical properties of the reconstructed
density field are contained in the growth function $\phi(z)$.
To study how lensing constrains this function, we begin with a 
model-independent approach.  

Consider the set of binned growth
rates $\phi_i$ as the parameters to be estimated.  
With fine-binning of the density field $\Delta z =0.1$, 
this leads to estimates that have
large correlated errors since the reconstruction effectively
takes differences of noisy data.  The long-time scale evolution of the density field is faithfully
preserved in the reconstruction \cite{HuKee02}.   Since structure grows in linear
theory on the expansion time scale in gravitational
instability models, this information is sufficient to constrain
cosmology.   

To better understand the information contained therein, consider
the principle component or eigenvector decomposition of 
the Fisher matrix ${\bf F}_{\phi\phi} =
{\bf S} \Lambda {\bf S}^t$
and the linear combinations of the data they define
\begin{eqnarray}
\esignal{\lambda}  =  {\bf S}^t \esignal{\phi}\,, %\nonumber\\
\qquad
\corr{\lambda\lambda}  =  \Lambda^{-1}\,.
\end{eqnarray}
In other words,
the eigenvectors are the redshift representation
of a new basis that is complete and yields uncorrelated, 
orthogonal measurements with variance 
given by the inverse eigenvalue $1/\lambda_i$.
The largest eigenvalues
correspond to the minimum variance directions and are shown in
the upper panel of Fig.~\ref{fig:eigengrow}.  
The first mode has a broad single peak between $0< z < z_{\rm med}=1$ 
corresponding to the  bell-shaped weight in the lensing projection of
Eqn.~(\ref{eqn:denproj}).  The higher modes exhibit oscillatory structure
and capture information on the low order derivatives of the
growth function around this intermediate redshift as well as the
region $z>z_{\rm med}$.
The spectrum of
eigenvalues $\rooteigen$, scaled to represent inverse rms noise and 
normalized for $\fsky=0.1$ are shown in the bottom panel.
Although the eigenvalues reflect only the noise properties and not the 
signal-to-noise, the growth rate in the
fiducial cosmology is sufficiently flat so that 
$[\signal{\lambda}]_i \rooteigen$ shares the same form (bottom panel).
Most of the signal comes from the first few 
eigenmodes but the signal-to-noise in the first 10-15 eigenmodes
remains substantial for a large survey.  

\begin{figure}[tb]
\centerline{\epsfxsize=3.5truein\epsffile{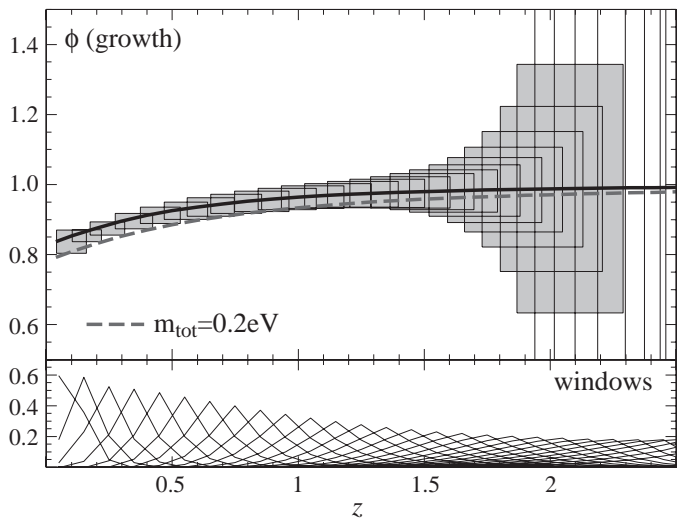}}
\caption{\footnotesize Localized constraints on the growth function $\phi=D_{\rm grow}/a$:
 (upper)
projected error boxes on the growth function with $1\sigma$
error bars on the Fisher-square root representation
with band widths taken from the window functions 
(lower).  For comparison the growth function for an $m_{\rm tot}=0.2$eV
model with three equal mass neutrinos is shown (dashed line).
}
\label{fig:decorrgrow}
\end{figure}

Although this principle component analysis is ideal for exposing
the nature of the information, the oscillating windows makes it somewhat difficult 
to visualize its impact for model testing.  For this purpose, it may be preferable
use more localized linear combinations that retain the uncorrelated
property at the expense of having overlapping (non-orthogonal) windows.  
Consider the linear
combinations defined by rows of  the ``square root" of the Fisher matrix \cite{HamTeg00}
\begin{equation}
{\bf G} \equiv {\bf S} \Lambda^{1/2} {\bf S}^{t}\,,
\label{eqn:sqrtfisher}
\end{equation}
and with a normalization chosen so that the window elements sum to unity.  
As shown in Fig.~\ref{fig:decorrgrow},
they yield well-localized windows and provide a visualization of
the data with error boxes
whose width is determined as that enclosing the central $60\%$ of
the window.  We show here the errors appropriate for a survey
with $\fsky = 0.1$.

Any model-independent constraints on the growth
rate can be translated into a model-dependent one by examining
the $\chi^2$ of the model fit.  In terms of Fisher forecasts, 
this is equivalent to a re-parameterization through the
transformation law (\ref{eqn:fishertrans}). 
As an example consider
a family of growth functions that represent a rescaling and
pivot around a fixed $z_p$ from the fiducial model
\begin{equation}
\phi(z) = A \left({ 1+z \over 1+z_p} \right)^{p} 
\phi_{\rm fid}(z) \,.
\end{equation}
The pivot point can be chosen to decorrelate the errors between
$A$ and $p$ by examining Fisher re-parameterization of $z_p$
\cite{EisHuTeg99b}.  This choice then has the nice property that the 
error $\sigma(A)$ in the two parameter model ($A,p$) 
are also those of the single parameter 
family of models ($A$).  For the fiducial model and survey
$z_p=0.64$, $\sigma(A) = 0.0023 \sifsky$, 
$\sigma(p) = 0.0089 \sifsky$.  Physically, such constraints
would limit the fraction $f$ of the dark matter
in a smooth component, for example a light neutrino below
its free-streaming scale. 
A smooth component induces a change in the growth
rate in the matter dominated epoch of $p =  3f/5$ and a consequent
change in the amplitude compared with the initial conditions
of $\delta A \approx 4 f$ (e.g. \cite{HuEisTeg98}).   

We
show a neutrino model with a total mass $m_\tot = 0.2$eV distributed
equally into three species \cite{BeaBel02} in Fig.~\ref{fig:decorrgrow}. 
This test is potentially substantially more powerful than its galaxy clustering
analogue due to the lack of an unknown bias \cite{HuEisTeg98,Elgetal02}.
While most of
the constraint would come from the amplitude $A$, information on 
the growth index $p$ is useful for distinguishing such effects
from those of the dark energy (\S \ref{sec:density}) and uncertainties
in the initial conditions (\S \ref{sec:discussion}). 
By varying the pixel size one can in principle test the scale 
dependence of growth rate predicted in such models.
Likewise, deviations would occur if the dark energy is not
effectively smooth on the scale of the pixels.  Lensing tomography 
offers a unique opportunity to test the clustering properties of both the dark energy and
the dark matter on intermediate scales.

\section{Dark Energy Density Evolution}
\label{sec:density}

If the distance-redshift relation $D(z)$ is uncertain due to the dark energy,
then these degrees of freedom must be incorporated into the statistical
forecasts as well.  For simplicity, we will in this context assume that the dark 
matter is composed solely of cold dark matter. 

Fortunately for a smooth dark energy component, both
the distance-redshift relation (\ref{eqn:distance}) and the 
growth rate are fixed by the dark
energy density evolution $\rho_{\de}(z)$ \cite{deprobes}.
Here the growth rate obeys
\begin{eqnarray}
\frac{d^2 \phi}{d \ln a^2}  &+&
\left[ \frac{5}{2} - \frac{3}{2} w(z) \Omega_{\de}(z) \right]
\frac{d \phi}{d \ln a}   \nonumber\\
&+& 
\frac{3}{2}[1-w(z)]\Omega_{\de}(z) \phi =0\,,
\end{eqnarray}
where the initial conditions are
$\phi=1$ and $d\phi/d\ln a=0$ and the equation of state
\begin{equation}
w(z) \equiv \frac{p_\de}{\rho_\de} = -\frac{1}{3} \frac{d \ln \rho_{\de}}{d\ln a} - 1\,.
\end{equation}

As in the case of the growth rate, we chose a model-independent
parameterization as the primary representation. Consider the dark energy
density in redshift bins \cite{WanGar01,Teg01}, specifically 
\begin{equation}
d_i = \ln \left( \frac{ \rho_{\de}(z_i)}{\rho_{{\rm cr}0}} \right)\,.
\end{equation}
We choose $\Delta z$ of
the bins to be the same as the density
reconstruction $\Delta z=0.1$. 
Since $\rho_{{\rm cr}0}$ is the critical density today, in the limit
of fine binning $d_1 = \ln \Omega_{\de}$.
We also require a finite  $w'(z) \equiv dw/dz$ to ensure that the 
dark energy remains smooth.  Therefore we choose $d(z)$
as a spline interpolation of $d_i$.

\begin{figure}[tb]
\centerline{\epsfxsize=3.5truein\epsffile{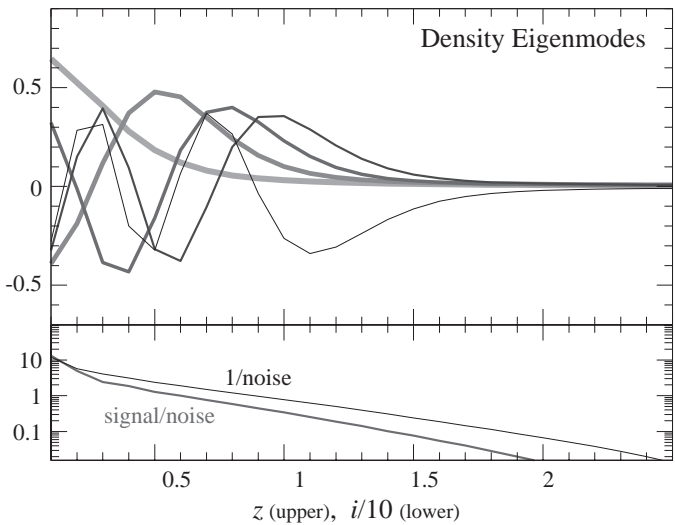}}
\caption{\footnotesize Principle components of the dark energy density $d \equiv
\ln(\rho_{\de}/\rho_{{\rm cr}0})$: (upper) first five
eigenfunctions of the growth function (thick to thin) plotted in 
redshift; (lower) rank ordered inverse rms noise $\rooteigen$ (thin) and 
signal-to-noise ratio in the fiducial model.  Higher eigenfunctions pick
up evolution in the dark energy density which is absent in the fiducial model.}
\label{fig:eigenden}
\end{figure}

Again, the recovery of  finely-binned parameters is noisy and correlated
across neighboring bins. The information content is best revealed
through the Fisher principle component analysis.
Shown in Fig.~\ref{fig:eigenden} (top) are the first 5 eigenfunctions.
The qualitative difference between the eigenfunctions of the growth and
that of the density is the presence of  substantial low redshift information in 
the latter.  The dark energy density at low redshift affects the distance
to all higher redshifts.   In Fig.~\ref{fig:eigenden} we show the eigenvalues
as $\rooteigen$ and the signal-to-noise ratio for the fiducial model and
$\fsky=0.1$.

\begin{figure}[tb]
\centerline{\epsfxsize=3.5truein\epsffile{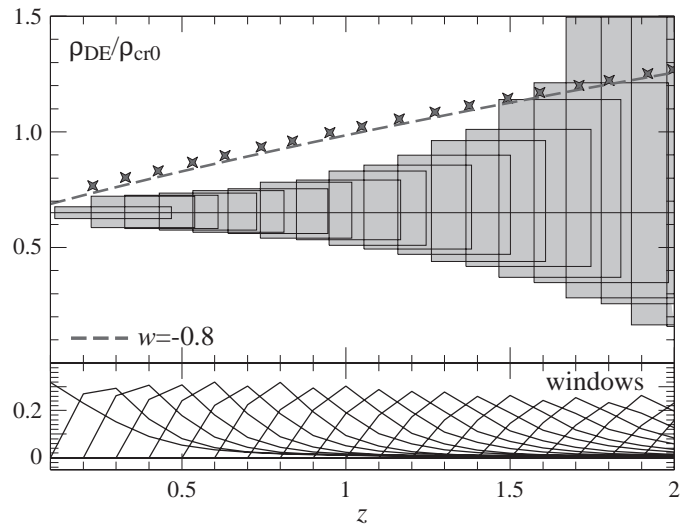}}
\caption{\footnotesize Localized constraints on the dark energy density: (upper)
projected error boxes on the growth function with $1\sigma$
error bars around the fiducial model (line) for 
the Cholesky representation where  band widths are 
taken from the window functions 
(lower).  
For comparison a model with $w=-0.8$ is shown both in the predictions
for the modes (points) and the density function itself (curve).  The small difference
reflects the non-locality of the windows. Any deviation from constancy represents
dark energy that is not a cosmological constant.}
\label{fig:decorrden}
\end{figure}

It is again desirable to find an uncorrelated but more localized representation 
of the data.  Unfortunately, the Fisher square-root
technique of Eqn.~(\ref{eqn:sqrtfisher}) does
not yield localized windows for the dark energy density.  Instead we
choose a close analogue, the Cholesky decorrelation \cite{HamTeg00} where 
the windows are the columns of ${\bf L}$ and ${\bf F}_{dd} = {\bf L}{\bf L}^t$
again normalized to sum to unity.  The windows are shown in Fig.~\ref{fig:decorrden} (lower
panel).  
The windows also have the interesting property that
they are strictly zero below some minimum redshift.

In Fig.~\ref{fig:decorrden} (upper panel)
we show the projected constraints on these localized modes 
compared with the predictions for a $w=-0.8$ model (points) and the
actual function $\rho_{\de}(z)$ (curve) in this model.   The deviation between the
points and the curve reflects the non-locality of the windows.  Note that
for a cosmological constant model ($\rho_{\de}=$ const.) there is no deviation by
definition (straight line) 
and that the expectation value of all points is $\ln \Omega_{\rm DE}$.   
Any statistically significant difference in the values of the points in this
reconstruction represent a detection
of a dark energy component that is {\it not} a cosmological constant \cite{WanGar01}.  This
is true in spite of the non-locality of the windows and independently
of the model-dependent parameterization of the dark energy.

Again one may always test specific models for the dark energy from
the model-independent parameterization.
Many dark energy models can be parameterized by 
$\Omega_{\de}$, $w(z_{\rm eff})$, and $w' = dw/dz |_{z_{\rm eff}}$. 
The pivot point $z_{\rm eff}$ can be chosen to be the best constrained redshift or 
``sweet spot" by decorrelating the errors in $w$ and $w'$.
As was the case for the growth rate, the resulting errors 
on $w(z_{\rm eff})$ are the same as in the case of a two parameter model
$(\Omega_{\de},w)$. 
For the fiducial model and survey, this is $z_{\rm eff}=0.33$ and 
$\sigma(w) = 0.016 \fsky^{-1/2}$.  Note that this constraint is marginalized
over $(\Omega_{\de},w')$ {\it without} prior assumptions to their values.  

\begin{figure}[tb]
\centerline{\epsfxsize=3.5truein\epsffile{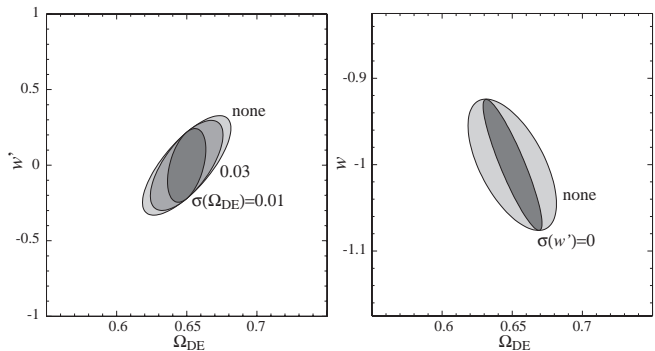}}
\caption{\footnotesize Error ellipses (68\% CL) for dark energy parameters ($\Omega_{\de}$, 
$w$, $w'$):  (upper) the ($\Omega_{\de},w'$) plane showing the degeneracy 
direction and the efficacy of external information on $\Omega_{\de}$;
(lower) the ($\Omega_{\de},w$) plane with complete and no external information on $w'$. 
Here $w$ is defined
as the equation of state at $z=0.33$.   
  The fiducial
model and survey are assumed throughout with $\fsky=0.1$.}
\label{fig:ellipse}
\end{figure}

As in the case of supernovae luminosity
distance measures \cite{WelAlb02,MaoBruMcMSte02}, there remains 
a degeneracy between $\Omega_{\de}$ and 
$w'$ in that they may both be adjusted upward to keep the dark energy
density at the well-constrained low redshifts fixed.  
In Fig.~\ref{fig:ellipse} shows, we show the
$68\%$ confidence region with various assumptions of prior knowledge on
$\Omega_{\de}$: none, $\sigma(\Omega_{\de})=0.03$, $=0.01$. 
With $\Omega_{\de}$ fixed, the errors become $\sigma(w')=0.046\fsky^{-1/2}$.
With no prior, the errors degrade to $\sigma(w')=0.069 \fsky^{-1/2}$.

Conversely
if $w'$ is fixed, the constraints on the dark energy density
improve to  $\sigma(\Omega_{\de}) = 0.003 \fsky^{-1/2}$.  This
two parameter family ($\Omega_{\de}, w$)   represents the amplitude and slope of the
dark energy density itself at a normalization point of $z=0$.  
The remaining degeneracy with $w$ can be removed by again going
to the 
``sweet spot", here $z=0.13$ where the errors on the dark energy
density improve by a factor of 1.5. 
Again this also represents the errors on $\Omega_{\de}$ in the single
parameter family of dark energy models $\sigma(\Omega_{\de}) = 0.002 \fsky^{-1/2}$. Errors on $w$ of course remain unchanged.

\section{Discussion}
\label{sec:discussion}

We have shown that lensing surveys covering more than a few percent of
the sky with good photometric redshift information can
probe the time evolution of the linear growth function and distance-redshift relation,
both of which are sensitive to properties of the dark energy and dark matter.
Specifically, we have tested a model-independent parameterization of
the linear growth rate and/or dark energy density discretized into bins in 
redshift.    Deviations in the growth rate would indicate a component of
the dark matter that is not effectively cold or dark energy that is not 
smooth on the lensing scale.  Deviations
in the constancy of the dark energy density would rule out a cosmological
constant model.

In this exploratory study, we have made several simplifying assumptions that would
need to be addressed in a concrete implementation. 
Perhaps the primary one is that future CMB measurements will completely fix the
high redshift cosmology.   The most uncertain piece involves the amplitude of
the initial fluctuations on the scales relevant to the lensing pixels,
$k \sim 0.05$ Mpc$^{-1}$ for degree scales.  
Fortunately, the pivot point of CMB anisotropy experiments with several
arcminute scale resolution is sufficiently close to the lensing scale 
that the slope and shape of the initial power spectrum do not 
cause much ambiguity \cite{Hu01c}.
% Planck has a pivot point of k=0.1Mpc^{-1} and sigma(\ln delta_zeta) = 0.0042

To fully utilize the lensing information, the initial amplitude must be fixed
to an accuracy better than the amplitude of the growth function, which we have
found to be  $\sim 0.002 \sifsky$, i.e. percent level accuracy for 
 surveys of several thousand square degrees.  
For CMB anisotropies, this precision requires that the optical depth 
during reionization must be determined to $\sigma(\tau) \sim 0.01$ to
resolve the amplitude degeneracy.
If determined from CMB observations alone, this will require polarization
measurements with a precision comparable to the  Planck satellite \cite{planck},
which can in principle achieve $\sigma(\ln \delta_\zeta)=0.0044$ at $k=0.05$ Mpc$^{-1}$
\cite{Hu01c}.  Direct measurements
of the reionization epoch can also resolve the ambiguity \cite{Hu01c}.
Even in the absence of this information, the {\it evolution} of the growth and
luminosity-distance relation are still constrained.   These issues are
best addressed through joint parameter estimation.

On the lensing side, the most important assumption is that the noise in
the convergence measurements is well-calibrated and not significantly larger
than the projections based on intrinsic ellipticities.  Even aside from the demanding
requirements for control of systematic errors, there may be intrinsic correlations
in the ellipticities \cite{intrinsic} that need to be modeled or avoided by increasing the pixel scale
and redshift bin widths.   The recovered information is largely insensitive to
the redshift bin width since the high signal-to-noise modes are all low frequency.
Errors scale roughly as $A_{\rm pix}^{1/2}$ due to the loss of independent modes
in a fixed survey area.

We have also neglected sample covariance between
the pixels but note that we have correspondingly neglected the information contained in such
correlations.  Indeed, we have completely neglected the information contained in the
non-linear regime which in fact contains the majority of the information from
lensing tomography \cite{HuKee02}.   Clearly, future studies will be required
to see how best to mine the model-independent information contained in lensing tomography.

{\it Acknowledgments:} I thank D. Huterer and C.R. Keeton
for useful conversations and E. Linder for pointing out
a typo in the draft.  This version includes errata from a coding bug, affecting constraints
involving the present dark energy density, pointed out by K. Abazajian and S. Dodelson.
 WH is supported by NASA NAG5-10840 and 
the DOE OJI program.  

%
% test larger pixels
%

\end{document}